\documentclass[12pt,a4paper,fleqn]{scrartcl}

\usepackage{graphicx} 
\usepackage{amsmath} 
\usepackage{amssymb} 
\usepackage{hyperref} 
\usepackage{authblk} 
\usepackage{natbib} 
\usepackage{afterpage}

\usepackage{framed}
\usepackage{lipsum}
\usepackage{verbatim}

\title{The Theory of Intrinsic Time}
\subtitle{A Primer}

\author{James B. Glattfelder}
\author{Richard B. Olsen}
\affil{Lykke Corp}
\date{\small \today}

\begin{document}

\maketitle


\begin{abstract}
    The concept of time mostly plays a subordinate role in finance and economics. The assumption is that time flows continuously and that time series data should be analyzed at regular, equidistant intervals. Nonetheless, already nearly 60 years ago, the concept of an event-based measure of time was first introduced. This paper expands on this theme by discussing the paradigm of intrinsic time, its origins, history, and modern applications. Departing from traditional, continuous measures of time, intrinsic time proposes an event-based, algorithmic framework that captures the dynamic and fluctuating nature of real-world phenomena more accurately. Unsuspected implications arise in general for complex systems and specifically for financial markets. For instance, novel structures and regularities are revealed, otherwise obscured by any analysis utilizing equidistant time intervals. Of particular interest is the emergence of a multiplicity of scaling laws, a hallmark signature of an underlying organizational principle in complex systems. Moreover, a central insight from this novel paradigm is the realization that universal time does not exist; instead, time is observer-dependent, shaped by the intrinsic activity unfolding within complex systems. This research opens up new avenues for economic modeling and forecasting, paving the way for a deeper understanding of the invisible forces that guide the evolution and emergence of market dynamics and financial systems. An exciting and rich landscape of possibilities emerges within the paradigm of intrinsic time. 
\end{abstract}

\section{Introduction}

Two main paradigms exist for decoding the workings of nature. One approach views its evolution as a smooth and continuous process. This thinking was instrumental in the inception of classical physics. Following the Pythagorean-Aristotelian tradition, mathematics was understood as the fundamental principle underlying the structure and function of the universe \citep[Ch. 2]{glattfelder2019information}. Specifically, a new brand of mathematics was divined that mirrored the smooth and continuous behavior thought to be the essential characteristic of the cosmos's dynamics \citep[Sec. 5.3.1]{glattfelder2019information}.

This analytical toolbox unleashed an astounding understanding of natural processes, conquering ever-new domains of application. To this day, the mathematical engine driving the most accurate physical theories---for instance, the standard model of particle physics and general relativity---depends on this framework. Physics Nobel Laureate Eugene Wigner mused about ``the unreasonable effectiveness of mathematics in the natural sciences" \citep{wigner1960unreasonable}. To this day, equations reign supreme in our fundamental understanding of reality, burrowing ever deeper into the fabric of existence.

In contrast, foundational to the second paradigm is the view that nature's evolution is rule-based, progressing in discrete steps. In other words, reality is computational. The conceptual focus shifts from an analytical perspective to an algorithmic one. This approach has been very successful in decoding complexity, otherwise a seemingly insurmoutable challenge in the equation-based paradigm. Echoing Wigner's epistemological surprise, the computer scientist, theoretical physicist, and entrepreneur Stephen Wolfram commented on a remarkable feature of complex systems. In essence, what appears as complex behavior from afar emerges from simple rules of local interaction \citep{wolfram2002new,wolfram2020project}---a second ``miracle" rendering the cosmos amenable to the cognitive and information-processing faculties of humans. Thus, complex systems are best understood by analyzing the structure of the interactions contained within them, a feat an equation-based approach falls short of.

Central to this paradigm is the multifaceted, intricate, and elusive concept of information. What was initially understood as a human concept has turned out to play a fundamental role in the fabric of reality. By design, information lies at the heart of computation. However, information processing is not exclusively related to human affairs. The enigmas of life and consciousness can be reframed in this context, offering new insights. Perhaps most remarkable, we find the notion of information at the foundations of reality. Information is a candidate for being an ontological primitive; that is to say, it could be one of the fundamental building blocks of reality itself, if not the only one \citep{glattfelder2019information,glattfelder2025sapient}.

In summary, the human mind has two main methods of comprehending the patterns appearing in the tapestry of reality. The analytical approach utilizes equations to uncover the knowledge comprising the edifice of physical science. Then, an algorithmic understanding of complexity gives us access to the simple structuring mechanism underlying it \citep{glattfelder2014r}.

\section{The Rise of Intrinsic Time}

Within this latter paradigm, we find the concept of intrinsic time. While the flow of physical time is traditionally understood as a continuous process, intrinsic time is an example of an event-based conception of time. In other words, it is algorithmically defined. In general, it should be noted that \citep[p. 1]{glattfelder2022bridging}:

\begin{quote}
    The flow of time is a central tenet in the subjective perception of reality. Human consciousness is eternally locked in the continuous transition between the past and the future, experienced as the moment of ``now.'' In stark contrast to the experiential familiarity of time, its ontological structure is obscure. From philosophy \citep{mctaggart1908unreality} to physics \citep[Ch. 10]{glattfelder2019information}, the nature of time has been debated for centuries; sometimes, its reality even wholly rejected \citep{connes1994neumann,barbour2001end}. The discovery of time's malleability \citep{einstein1905elektrodynamik} and its resistance to being quantized \citep{dewitt1967quantum} only add to the enigma.
\end{quote}

Historically, the understanding of time has always been linked to the unfolding of physical events: from lunar calendars, sundials, hourglasses, and quartz clocks to atomic clocks, the passage of time is made visible by measuring physical changes. In essence, time can be viewed as a property arising from complex networks of events and their interactions governed by quantum mechanics and thermodynamics \citep{rovelli2019order}.

In economics, the notion of time usually plays a marginal role. Nonetheless, the practice of using an alternatively defined conception of time has a long tradition. Specifically \citep[p. 2]{glattfelder2022bridging}:

\begin{quote}
The idea of modeling financial time series in a new temporal paradigm goes back to
\citep{mandelbrot1967distribution} and has been a reoccurring theme
since \citep{clark1973subordinated,ane2000order,easley2012volume}. In essence, physical time is substituted with an event-based notion of time. This is to say that these novel measures of time are operationally defined using certain intrinsic features of the data being analyzed---as an example, driven by transaction numbers or trading volumes.
\end{quote}

Then, thirty years after the original inception of event-based time, a seminal paper from a team at Olsen \& Associates proposed a simple algorithm focusing on a key characteristic of financial time series: changes in the direction of the price evolution \citep{guillaume1997bird}. The contours of what is today understood as intrinsic time became visible. Historically, researchers at Olsen \& Associates had coined the term one year earlier \citep[p. 213]{dacorogna1996changing}:
\begin{quote}
The only information needed to define the [time] scale are the values of the time series themselves. Thus we have chosen to call this time scale \textit{intrinsic time}. The consequence of using such a scale is to expand periods of high volatility and concentrate those of low volatility, thus better capturing the relative importance of events in the market.
\end{quote}
However, although the methodology did allow for a decoupling from physical time by focusing on the intrinsic behavior of time series, it was defined analytically. The power of simple rules could only be fully harnessed within the directional change methodology unveiled in 1997.

The algorithmic formulation of intrinsic time functions similarly to its analytical counterpart: In periods of low market activity, the clocks tick slower and, conversely, speed up during phases of high market activity. The endogenous atoms of intrinsic time are given by the directional changes. In the simplest terms, they are reversals of price moves measured from local extrema at different scales (the details are given below). Directional changes are more sophisticated versions of what are known as drawups and drawdowns \citep{pospisil2009formulas}. The concept of directional changes has also entered computational finance, especially in the context of high-frequency data \citep{aloud2012directional,li2022measuring,tsang2024nowcasting}.

From a conceptual perspective, the development of intrinsic time reflects the acceptance that an abstract universal time does not exist and that the concept of time is always observer-dependent. In other words, the intuitions from special relativity are applied in a broader context. Generically, intrinsic time can be understood as a process of how time emerges within a system of interacting agents \citep{olsen1983interaction}. This idea marks a pronounced metaphysical shift, moving from the concept of time as an ontological category of the universe to understanding it as a secondary, derived property dependent on intrinsic interactions. It is interesting to note that a modern understanding of quantum gravity supports the idea that time (specifically spacetime) is an emergent property, perhaps derived from entangled quantum information, succinctly encapsulated by the mnemonic term ``ER=EPR" \citep{maldacena2013cool}.

In 2011, the concept of intrinsic time was further refined. Next to the event-based triggers described by the directional changes, a deeper layer of intrinsic activity was unveiled, called overshoots. Overshoots are robust statistical properties hidden in time series (see the discussion in the next section). In general, after every directional change, the price evolution continues, on average, one overshoot length before the following directional change disrupts it.

Intrinsic time is algorithmically derived from a selected threshold $\delta$, describing a price move in percent. The oscillatory nature of directional changes requires two modes to be specified: \textit{up} and \textit{down}. 
Without loss of generality, one can begin in an \textit{up} mode at a price of $x_0$ and set the extremum price to this value, $x^{ext} = x_0$. The directional change algorithm evolves as follows:
\begin{itemize}
\item if the price moves up, $x^{ext}$ is updated to this new value;
\item if the price moves down, the percentage difference between $x^{ext}$ and the current price is evaluated.
\end{itemize}
The inevitable outcome is that there will come a price move for which the percentage difference between $x^{ext}$ and the current price $x_t$ exceeds $\delta$. A directional change tick is registered, the mode changes to \textit{down}, and $x^{ext} = x_t$. Now, the algorithm continues correspondingly, with switched directions (i.e., $x^{ext}$ is updated on lower prices, and $\delta$ measures its difference to higher prices). Also, note that two intrinsic time clocks will be synchronized after two directional changes even if they were initialized in different modes.
The pseudocode for the algorithm can be found in \citep{glattfelder2011patterns}.

\begin{figure}[p]
 \centering
 \includegraphics[width=0.9\linewidth]{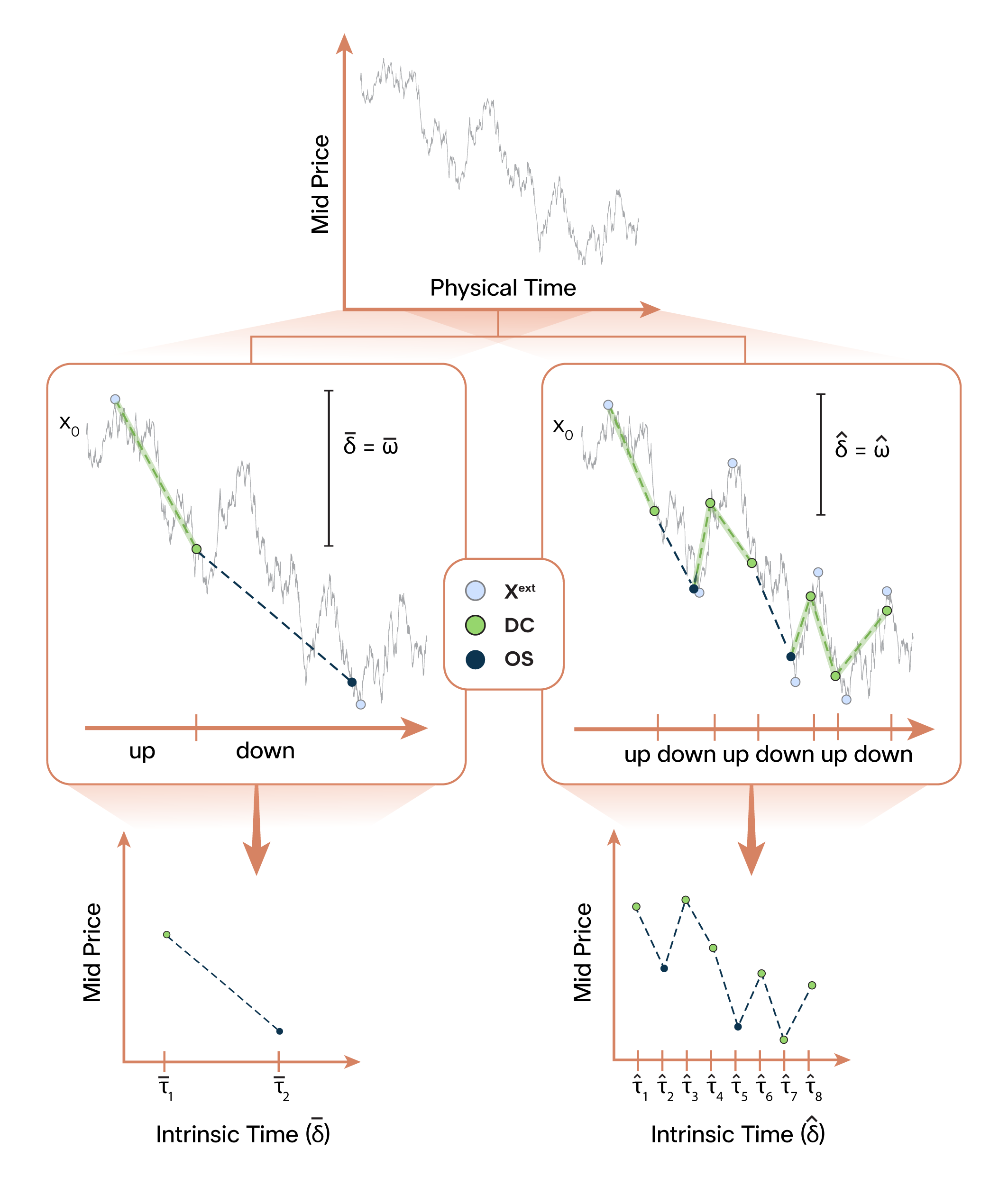}
 \caption{Intrinsic time transformations: The raw time series is decomposed into directional change (DC) and overshoot (OS) events according to the procedure defined in the main text utilizing two thresholds for measuring the intrinsic time events. Specifically, $\bar \delta$ and $\hat \delta$ define the directional changes and $\bar \omega$ and $\hat \omega$ the overshoots, with $\bar \delta = \bar \omega$ and $\hat \delta = \hat \omega$, respectively. The equality of the variables is derived in Eq. (\ref{eq:os}). In effect, intrinsic time defined by different thresholds of resolution yields transformations from a tick-by-tick representation of time series to event-based price curves, evolving in intrinsic time increments $\bar \tau_i$ and $\hat \tau_j$, respectively. The dependency on physical time is thus substituted by a dependency on intrinsic market activity, defined at chosen levels of granularity, revealing the underlying coastlines.}
 \label{fig:osdc}
\end{figure}

If the price continues to move by $\delta$ in the same direction after a directional change, the first overshoot is registered, representing the next tick of intrinsic time. Now, if the price evolution continues in the same direction by $\delta$, a second overshoot is detected, and so forth. It is thus possible to define the coastline of a time-series as the sum of all price moves of $\delta$ belonging to directional changes or overshoots. Crucially, intrinsic time is a multi-scale concept, defined by a set of thresholds $[\delta_1, \dots, \delta_n]$. As a result, for every smaller choice of $\delta$, a more granular and longer coastline emerges. In essence, every directional change threshold allows the tick-by-tick price curve to be transformed into a representation filtered by intrinsic market activity. See Fig.  \ref{fig:osdc} for a visual representation.

\afterpage{\clearpage}

The rigid application of physical time---i.e., the traditional approach to modeling financial time series assuming time intervals to be uniformly spaced, for instance, in seconds or days---is limiting and may obscure essential characteristics of financial time series. In contrast, the adaptive quality of an event-based conception of time offers a more flexible method for exploring the complexity and heterogeneity of market data. Within the new dynamic paradigm of intrinsic time, novel structures and regularities are uncovered, otherwise invisible to the inflexible cadence given by equidistant time intervals.

\section{The Power of Scaling Laws}

Viewed through the lens of the philosophy of science, scaling laws can be understood as laws of nature applicable to complex systems. Specifically \citep[p. 86]{glattfelder2025sapient}:

\begin{quote}
Laws of nature are regularities and structures in a complex
universe that critically only depend on a small set of conditions
and are independent of many other factors which could
potentially have an effect. Universal scaling laws can be
understood as such laws relating to nature’s complexity.
A scaling law is a simple polynomial relationship between
a function’s input and output. Specifically, a relative change
in the input quantity results in a proportional change in the
output quantity, independent of the initial size of the input.
Scaling laws are scale-invariant, meaning there is no preferred
or defining scale, reminiscent of the self-similarity of fractals.
Historically, the first study of scaling laws and scaling effects
can be traced back to Galileo, who investigated how different
attributes of ships and animals scale with size. Over 250
years later, the linguist George Kingsley Zipf studied rank-
frequency distributions in the 1950s. He discovered that in
English, the most common words “the,” “of,” and “and” are
used disproportionally. In contrast, all other words follow a
scaling-law relation of lesser and lesser importance. In 1964,
the economist and sociologist Vilfredo Pareto analyzed the
distribution of wealth and coined the term 80-20 rule. Also
known as the Pareto principle, this approximate rule says that
20\% of the population controls 80\% of the wealth. Formally, the
distribution Pareto identified was a scaling law.
\end{quote}

The hallmark of a scaling law is its linear behavior in a log-log plot. However, stringent statistical tests should be performed to rule out misidentification \citep{clauset2009power}. Scaling-law relations can be observed in a spectacular diversity of complex systems. Generally, there are four basic types of scaling laws to be distinguished: (1) allometric scaling laws, (2) scaling-law distributions, (3) scale-free networks, and (4) cumulative scaling-law relations \citep[Sec. 6.4.2]{glattfelder2019information}.

The first category bridges biology, ecology, and physics, describing how different biological quantities scale in relation to each other as the size of an organism changes. For example, a relationship exists between a mammal's body mass and either the number of heartbeats or the lifespan. The larger the animal, the slower its heart beats, but the longer it lives. As a remarkable consequence, a fundamental invariant of life emerges: the number of heartbeats of any mammal is constant over its lifetime and is approximately $1.5 \cdot 10^9$ \citep{west1999origin}.

The second category spans a vast array of phenomena, as the dynamics of many complex systems are not captured by a Gaussian distribution characterized by a mean scale. In a nutshell, scaling-law distributions reveal how a small fraction of a population has a dominant influence while the remaining population is mostly inconsequential. They emerge in such diverse phenomena as \citep{newman2005power}:
\begin{itemize}
\item the size of cities, earthquakes, moon craters, solar flares,
computer files, sand particles, wars, and price moves in
financial markets;
\item the number of scientific papers written, citations received
by publications, hits on web pages, and species in
biological taxa;
\item the sales of music, books, and other commodities;
\item the population of cities;
\item the income of people;
\item the frequency of words used in human languages and of
occurrences of personal names;
\item the areas burnt in forest fires.
\end{itemize}

The third category describes the organizing principles seen in real-world complex networks. The turn of the millennium brought about a revolution in the fundamental understanding of their structure and dynamics. Seminal discoveries were the small-world \citep{watts1998collective} and scale-free \citep{barabasi1999emergence} nature of these networks, bringing about a new science of networks \citep{dorogovtsev2002evolution}. In detail, scale-free networks have a degree distribution following a scaling law. Again, the scaling relation dictates that a few super-hubs are of high relevance while most nodes in the network play a subordinate role.

Finally, the last type of scaling-law relation manifests in collections of random variables known as stochastic processes. Notable examples are financial time series, where empirical scaling laws govern the relationship between various measurable quantities. Interest in the scaling relations inherent in market data was sparked in 1990 by a seminal paper by researchers at Olsen \& Associates relating the mean absolute change of the logarithmic mid-prices, sampled at different time intervals over a sample of size, to the size of the time interval \citep{muller1990statistical}. Later, in 1997, the group discovered a second scaling law relating the number of directional changes $N$ in a sample to the directional-change threshold $\delta$ \citep{guillaume1997bird}:
\begin{equation}
N(\delta) = a \delta^b,
\end{equation}
where $a$ is a proportionality constant and $b$ is the scaling law exponent. Then, in 2011, 12 independent new empirical scaling laws were discovered in foreign exchange data, holding for close to three orders of magnitude and across 13 currency exchange rates \citep{glattfelder2011patterns}. Similarly to the scaling relation of 1997, this new set of laws emerged within the directional change framework. One such scaling relation proved to be remarkably insightful. The size of the average overshoot length $\omega$ after a directional change is approximately the size of the threshold $\delta$:
\begin{equation}
\left< \omega(\delta) \right> \approx \delta, \label{eq:os}
\end{equation}
where the angle brackets denote the sample average. Thus, the motivation arose for defining an overshoot increment to be the size of the directional change threshold, $\omega=\delta$, expanding the scope of the intrinsic time framework. It should be noted that Eq.(\ref{eq:os}) only holds for the average behavior. The individual instances of actually measured overshoot lengths reveal that the average value is never actually manifested. There is a large variability in the distribution where most occurring overshoots are small, with some disproportionately large ones conspiring to yield the average behavior seen in Eq.(\ref{eq:os}).

Herein lies the power of scaling laws. They uncover an underlying organizing principle coordinating over many orders of magnitude. It is surprising to find such regularity in the structure and evolution of complex systems. Once detected, these patterns make them amenable to systematic analysis and predictive modeling. This allows researchers to discern novel behaviors and relationships that provide insights into the fundamental dynamics of complex systems across various scales and multiple domains.

\section{Decoding Complexity}

The insights from the intrinsic time framework inspired further research. For instance, the concept of multi-scale liquidity was introduced \citep{golub2016multi}, systematic trading strategies devised \citep{golub2018alpha}, a variation of the notion of volatility defined \citep{petrov2019instantaneous}, and an agent-based framework formulated \citep{petrov2020agent}. The notion of intrinsic time can be extended to a multi-dimensional methodology, incorporating more than one financial time series \citep{petrov2019intrinsic}. Other research groups have also adopted the directional-change framework \citep{tsang2017profiling,tsang2022directional,tsang2024nowcasting}.

Recent work uncovered an additional scaling law relating to the behavior of overshoots. A central result was a recipe for the decomposition of time series into their liquidity and volatility components, both only visible through the lens of intrinsic time \citep{glattfelder2022bridging}:
\begin{equation}
\label{eq:gap}
  \langle r(\Delta t)\rangle_2 \propto \langle \omega(\delta)-\delta\rangle_2   N(\delta),
\end{equation}
where $r(\Delta t)$ denotes the price changes or returns sampled in physical time, and $\langle x \rangle _2= \frac{1}{n}\sum_{i=1}^n x_i^2$ is the sample average of the squared values. Note that $\langle \omega(\delta)-\delta\rangle_2$ represents the  variability of overshoots (and scales with $\delta^2$). Then, by observing that the overshoots function as a proxy for liquidity \citep{golub2016multi} and the number of directional changes measures volatility \citep{petrov2019instantaneous}, financial time series can be non-trivially decomposed into these two defining characteristics, yielding novel understanding.

Taking a step back, the utility of intrinsic time can also be assessed in more general terms. Humanity's technological prowess is truly breathtaking, witnessed by unprecedented leaps in progression, overshadowing former advancements thought to be unsurpassable. Fuelling this cognitive revolution is the knowledge generated in the analytical paradigm, epitomized by physical science. In contrast, our understanding of complexity is still in its infancy. Although complexity science emerged as a fusion of various intellectual traditions---including cybernetics, systems theory, early artificial intelligence, cellular automata, non-equilibrium thermodynamics, agent-based modeling, non-linear dynamics, fractal geometry, chaos theory, network science, metacybernetics, and complex systems theory---we are still waiting for deeper insights into emergent phenomena like the enigmas of life and consciousness \citep{glattfelder2025sapient}.

Perhaps the most glaring failure regarding our understanding of complexity relates to finance and economics. It is bitterly ironic that the very systems humans collectively engineered, representing the backbone of global stability and progress, are also among the least understood in terms of their intricate workings, long-term sustainability, and susceptibility to systemic risk \citep[Ch. 7]{glattfelder2019information}. It is within this domain that we have an excellent opportunity to apply some general insights gained from intrinsic time. Specifically, the dissemination of information triggering the responses of economic agents, mythologized by Adam Smith's ``invisible hand" \citep{smith1776inquiry}, appears in a new light. By accounting for the observer-dependent reference frames defined by an event-based framework, it is naive to assume the efficient collective allocation of resources in response to signals, even by discounting for human irrationality \citep{kahneman2003maps}. Accepting the high degree of idiosyncrasy, variability, and heterogeneity in the individual perception and processing of information---exemplified by the ticking of local intrinsic clocks---collective behaviors like herding and cascading effects, so detrimental to stability and sustainability, can be reevaluated in a framework more attuned to the nuances of complex systems of interacting agents.

The exploration of the intrinsic time framework, galvanized by the emergence of scaling laws, represents a paradigm shift in the overall understanding of financial markets and complex systems. It underscores a profound departure from static, linear models of analysis towards dynamic, event-driven frameworks that more accurately reflect the complexities of real-world phenomena. This shift fundamentally challenges the traditional perceptions of time and uncovers its event-based emergence in the context of an interaction-based, algorithmic worldview. A rich landscape for innovation in both theoretical and practical terms is discovered. Essentially, a more subtle grasp of nature's underlying dynamics becomes possible within this paradigm. Thus, intrinsic time seems like an indispensable item in an intellectual toolbox designed to decode the complexity arising from humanity's expanding quest for knowledge.


\bibliographystyle{apalike} 
\bibliography{my} 

\end{document}